Charge transport in thin-film transistors based on liquid-crystalline phthalocyanines

Lindiomar. B. Avila*[1,2], Zuchong Yang[2], Ilknur H. Eryilmaz[2], Lilian Skokan[2], Leonardo N. Furini[1], Andreas Ruediger[2], Harald Bock[3], Ivan H. Bechtold[1], Emanuele Orgiu*[2]

[1] Departamento de Física, Universidade Federal de Santa Catarina−UFSC, 88040-900 Florianópolis, SC, Brazil

[2] Institut national de la recherche scientifique, Centre Énergie Matériaux Télécommunications, 1650 Blvd. Lionel-Boulet, Varennes, QC, Canada

[3] Centre de Recherche Paul Pascal, Université de Bordeaux & CNRS, 115 Avenue Schweitzer, 33600 Pessac, France.

## Abstract Figure

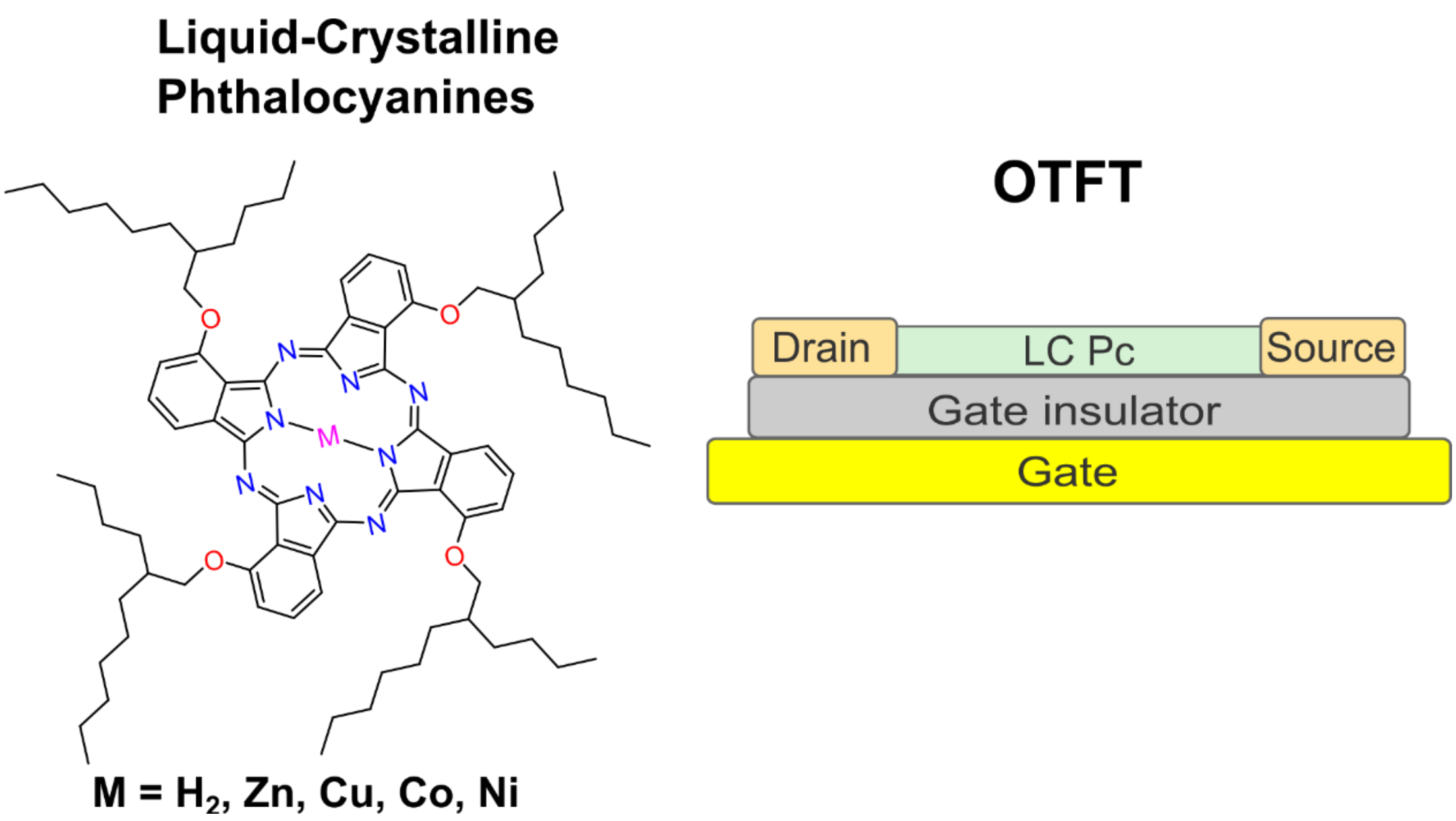

Corresponding author email: lindiomarbaj@gmail.com, emanuele.orgiu@inrs.ca

**Abstract:**

We investigate a series of liquid-crystalline phthalocyanines (metal-free and Cu, Zn, Ni, Co complexes) and their electronic performance as active layers in organic thin-film transistors (OTFTs). Raman spectroscopy reveals metal-dependent distortions of the phthalocyanine macrocycles, reflected in systematic shifts of the $C-N-C$ and $M-N$ vibrational modes. When integrated into OTFTs, all compounds exhibited markedly enhanced current response under ultrahigh vacuum compared to an $N_2$- atmosphere, demonstrating that intrinsic charge transport is suppressed by atmospheric species. Temperature-dependent measurements (in the $100-300\,K$ range) revealed that the threshold-voltage shifts in the devices were driven by deep interface and bulk traps, while all devices displayed thermally activated mobility with low activation energies ($\approx 14-20\,meV$). These results highlight how mesomorphic order, metal coordination, and environmental conditions collectively govern charge transport in liquid-crystalline phthalocyanines, offering design guidelines for their use as self-assembled semiconducting materials in organic electronics. Our results therefore represent a systematic characterization of liquid-crystalline phthalocyanine-based OTFTs, revealing low trap-barrier energies and efficient charge-hopping pathways present in these materials.

## 1. Introduction

Phthalocyanines (PCs) have played a pivotal role in materials science since their discovery in 1907, initially attracting attention as robust dyes and pigments. Besides their strong absorption across the visible–near-IR spectrum, their remarkable photo- and thermal stability and ability to form extended $\pi - \pi$ stacked structures soon expanded their relevance far beyond their use as dyes. By the mid-20$^{th}$ century, PCs have emerged as a promising molecular platform for optoelectronic organic-based devices. A major milestone occurred in 1982, when Piechocki and Simon reported the first PC capable of forming a columnar liquid-crystalline mesophase, a discovery that opened new avenues for controlling supramolecular ordering in optoelectronic devices. Follow-up studies demonstrated pronounced charge-transport anisotropy in the columnar LC mesophases of peripherally substituted PCs, making them especially attractive for photovoltaics, field-effect transistors, and other applications requiring directional charge transport [1],[2], [3], [4].

Their chemical and thermal robustness, versatile coordination chemistry, and tunable electronic structures make PCs highly appealing for more device-oriented applications [5] [6]. Furthermore, peripheral functionalization of the phthalocyanine macrocycle has emerged as a powerful means to modulate intermolecular interactions and solubility, thereby enabling controlled solution processing without compromising the intrinsic electronic properties of the conjugated core [7]. Liquid-crystalline PCs have attracted significant attention because their mesophase behavior enables controlled molecular alignment, which minimizes structural disorder and facilitates directional charge transport along ordered molecular stacks, resulting in enhanced charge mobility and device efficiency [8] [9] [10] [11] [12]. Comparative studies of structural motifs, mesophase types, and alignment techniques continue to reveal how subtle molecular design choices can tailor the optoelectronic performance of PC-based materials [13].

All compounds in this work exhibit liquid-crystalline behavior, a property investigated comprehensively in a previous work [14]. It was shown that all the compounds encompassed within this study do develop columnar mesophases at temperatures below 300 °C. At room temperature, all phthalocyanine derivatives investigated exhibit a hexagonal columnar mesophase. While the metal-free $H_2$ homologue shows an uncommon hexagonal columnar arrangement characterized by a twelve-column unit cell, the $Zn, Cu, Ni$, and $Co$ analogues also stabilize hexagonal columnar order, with differences in lattice organization and clearing temperatures reflecting variations in metal coordination geometry and inter-disk interactions.

In the present study, liquid-crystalline phthalocyanines are integrated into organic thin-film transistors (OTFTs) to systematically correlate the mesomorphic order, metal coordination, and thermally accessed phase behavior with charge-transport properties under controlled environmental conditions. Beyond the structural and thermal properties previously reported, we investigate their electrical behavior under two distinct measurement conditions: under high-vacuum and an $N_2$-atmosphere. Moreover, by probing their transport characteristics over a wide temperature range, including low-temperature operation, we gain insight into the role of columnar ordering, inter-disk interactions, and trap dynamics in determining charge-carrier mobility.

## 2. Experimental

Phthalocyanines were used as active layers in bottom-contact, bottom-gate OTFTs, fabricated by spin-coating onto Si/$SiO_2$ substrates with gold source and drain electrodes. The solutions were prepared by dissolving Pc in toluene at a concentration of 10 $mg/mL$.

For the device fabrication, $100\ \mu L$ of solution were deposited to the substrate, followed by spin-coating at $1000$ rpm for $30$ seconds, a procedure that reproducibly yielded uniform thin films with a thickness of approximately $200$ nm. Initial electrical measurements were conducted in nitrogen atmosphere (glove box with $H_2O$ and $O_2$ levels $< 0.1$ ppm). Devices were left in the glove box overnight before the measurement to ensure the removal of oxygen and moisture from the surface. For low-temperature characterization, the devices were measured by means of a DynaCool PPMS (Quantum Design). The source, drain, and gate electrodes were connected to the sample puck using indium contacts, and the measurements were performed under high vacuum conditions ($P < 10^{-5}\, Torr$) using a Keithley 2636 source meter to apply voltage and record current.

### 2.1 *Materials preparation*

The synthesis and chemical characterization of the liquid crystalline PCs encompassed in this study was previously reported [13]; for the sake of completeness, a recap with the main experimental synthesis steps are also provided in the SI. Figure 1 shows the molecular structure of the different PCs, where M = $H_2$, Zn, Cu, Co and Ni. For simplicity, these fourfold ether-substituted PC derivatives are abbreviated in the following as $H_2$Pc, ZnPc, CuPc, CoPc and NiPc without explicit reference to the ether chains.

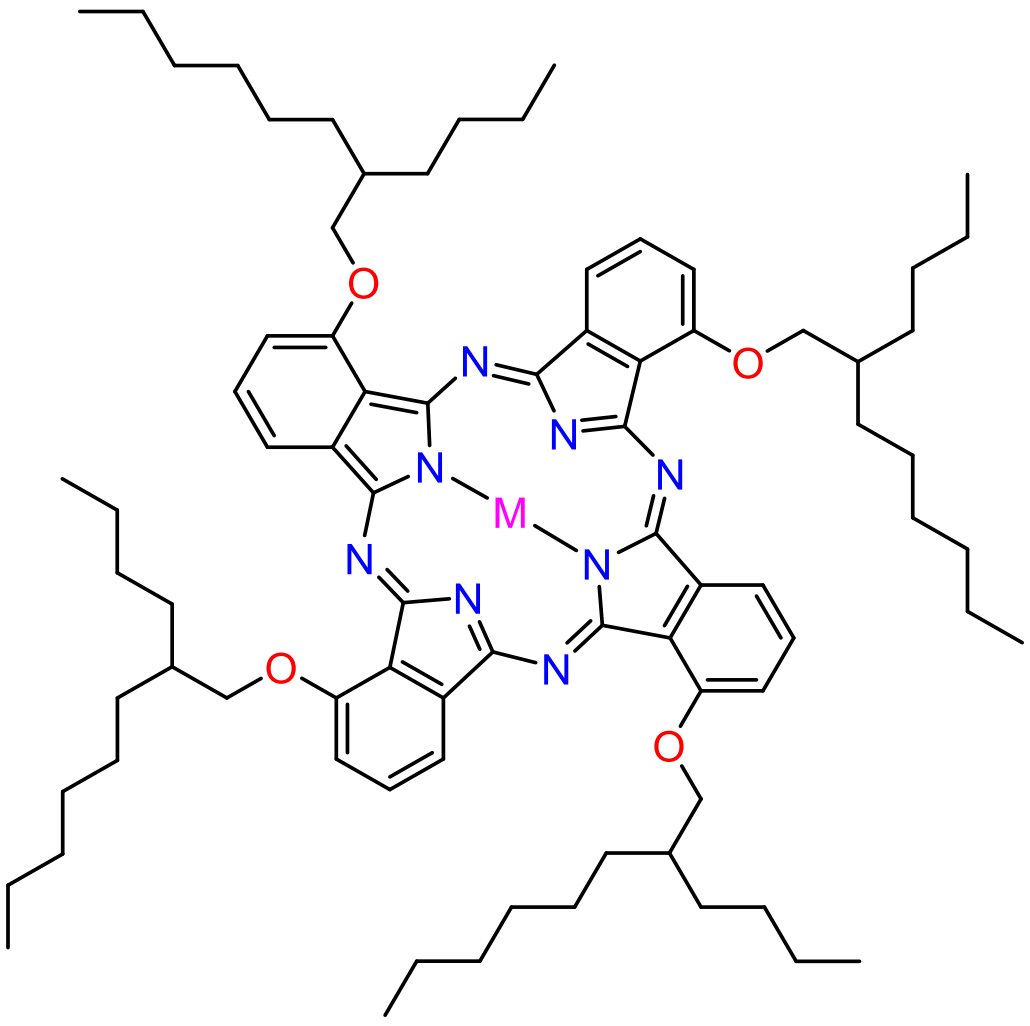

Figure 1. Chemical structure of the investigated Pc-based liquid crystalline compounds. M = Zn, Cu, Co and Ni. For metal-free Pc case, M = $H_2$

Homogeneous PC films were deposited by spin-coating on a $SiO_2$/highly doped $Si(p^{++})$ substrate. Atomic force microscopy (AFM) images before and after annealing, as shown in Figure S1, revealed that all PC films exhibited root mean square roughness ($R_{rms}$), values ranging from 1.1 to 1.6 nm, consistent with homogeneous surface morphology and uniform film formation under the employed fabrication conditions. Notably, after annealing, the metal-atom-containing PC (MPc) samples showed a decrease in $R_{rms}$, while the $H_2Pc$ sample exhibited a 30% increase in $R_{rms}$. The low surface roughness observed in this work contrasts with previously reported values, often an order of magnitude higher and further increasing upon annealing, despite all films being deposited under comparable processing conditions (identical solvent, solution concentration, and spin-coating parameters) and with similar thicknesses of approximately 200 nm [15], [13]. This suggests that depositing PC films onto smooth thermally-grown $SiO_2$ surfaces leads to particularly smooth and well-annealable films. This finding holds significant potential for applications in organic semiconductor devices such as OFETs, organic photovoltaic cells (OPVs), and organic light-emitting diodes (OLEDs), where surface smoothness and film uniformity are essential to optimizing device performance.

## 3. Results and discussion

### 3.1 ***Raman Characterization***

Raman spectra were collected for all compounds in powder form, and the corresponding results are presented in Figure 2. In general, all the metallic Pc compounds presented a sizeable signal-to-noise ratio, except for $CoPc$ where a quite modest intensity was observed. Although great caution should be taken when working with absolute Raman intensities, a self-absorption process due to the low Raman intensity for $CoPc$ using this laser line could be put forward [17], [15]. In essence, the Raman spectra of PCs are dominated by isoindole, pyrrole and macrocycle vibrational modes. $H_2$Pc's Raman spectrum presents intense band at 1397 $cm^{-1}$ assigned to C-C stretching from pyrrole moieties, 1504 $cm^{-1}$ attributed to -C-N- bridges between isoindole groups and 1601 $cm^{-1}$ assigned to conjugated carbons (-C=C-) and (-C=N-) groups [19] [15] [20]. The macrocycle deformation can be observed at 727 $cm^{-1}$ and the C-H deformation at 1095 $cm^{-1}$ [19] [21] [22]. The aforementioned bands are highlighted by stars in Figure 2. It was not possible to identify vibrational modes from the lateral alkyl chains, which confer the liquid-crystallinity. Usually C-H deformations appear around 2900 $cm^{-1}$, but in our case, no such band was observed. Bands associated with tertiary carbons (which are usually weak) or with ether groups were not observed either [21]. Nevertheless, one can speculate that the isoindole ring deformation can be affected by the presence of lateral chains. Such a vibrational isoindole ring deformation mode has been reported at 230 $cm^{-1}$ [19], and in our case the band was seen at 222 $cm^{-1}$.

The presence of metal atoms induces a significant change in the Raman spectra, mainly to the $-C-N-C$ isoindole bridges, which shifted from 1504 $cm^{-1}$ in $H_2$Pc to 1530 $cm^{-1}$ in NiPc, 1527 $cm^{-1}$ in CoPc, 1513 $cm^{-1}$ in CuPc and 1491 $cm^{-1}$ in ZnPc. The major shift was observed for NiPc (26 $cm^{-1}$), followed by CoPc (23 $cm^{-1}$), ZnPc (13 $cm^{-1}$) and CuPc (9 $cm^{-1}$). Whereas NiPc, CoPc and CuPc presented a shift to higher wavenumber, ZnPc showed a shift to lower wavenumber compared to $H_2Pc$. Such effects are related to the metal ion size and cavity diameter [17]. Tackley *et al.* [18] report different cavity diameters for metallic Pc correlating with their Raman spectra. Their findings reveal that the nickel derivative has a particularly small inner cavity. Thus, the isoindole moieties are closer to the nickel atom, changing the $-C-N-C-$ bridges and causing a large displacement (26 $cm^{-1}$) observed in the Raman spectrum (Figure 2). As the size of the metal ion increases, it fits in the cavity with less shift in the Raman spectra as observed for Co followed by Cu. [18]. In the case of Zn, the metal ion size is a little bigger than the cavity and the energy-minimised structure calculated by density functional theory (DFT) was obtained with the Zn atom lightly out of the phthalocyanine ring [20]. Such an effect can account for the shift toward lower wavenumbers observed in the Raman spectra. In addition, the ZnPc spectrum exhibits two intense bands at 1399 $cm^{-1}$, assigned to C–N stretching of the pyrrole units, and at 1595 $cm^{-1}$, attributed to C–C stretching within the isoindole rings [23] Metal–nitrogen vibrational modes are observed at 242 $cm^{-1}$ for CuPc, 245 $cm^{-1}$ for NiPc and ZnPc, and 246 $cm^{-1}$ for CoPc [24]. The identical frequencies observed for NiPc and ZnPc indicate similar effective metal–nitrogen force constants, likely arising from comparable coordination environments within the phthalocyanine macrocycle, with possible compensation between reduced mass and bond strength effects. Complementarily, macrocyclic vibrations (in-plane breathing and deformations) are observed in the 600-800 $cm^{-1}$ frequency range for all PCs [23]. The C-H deformations from benzene rings could be expected to be between 1000 and 1300 $cm^{-1}$ with emphasis on the strongest peak at 1095 $cm^{-1}$ for ZnPc.

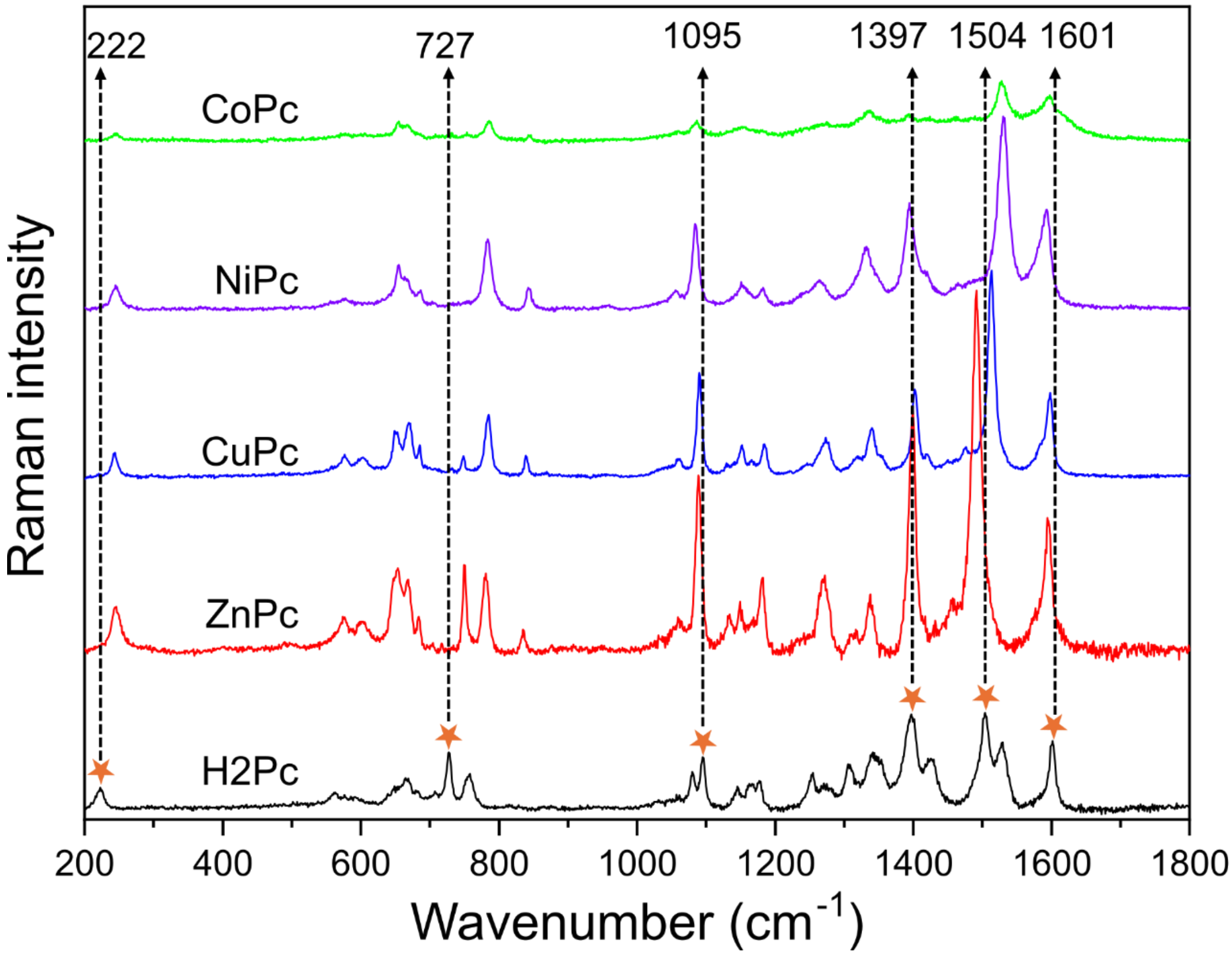

Figure 2. Raman spectra of metal-free phthalocyanine ($H_2Pc$) and its metal-coordinated derivatives (CuPc, ZnPc, NiPc, and CoPc)

### 3.2 ***Electrical Characterization***

Figure 3 shows the transfer characteristics of the five investigated molecules measured under two conditions: high vacuum inside a cryogenic semiconductor characterization system (PPMS DynaCool by Quantum Design) system and in nitrogen atmosphere room-temperature glovebox environment. In vacuum, MPcs are effectively isolated from atmospheric species, enabling their intrinsic semiconducting properties to dominate the device behavior. The enhanced current observed under vacuum conditions is likely associated with the reduced adsorption of environmental species, including residual $O_2$ and $H_2O$ present in the glovebox atmosphere, which are known to introduce charge-trapping states in organic semiconductors. Among the compounds studied, the metal-free $H_2Pc$ displayed the weakest electrical response, whereas all MPc-based devices exhibit drain currents within the same order of magnitude (Figure S2). The detailed transfer characteristics measured under nitrogen and vacuum, including their temperature dependence, are provided in the Supplementary Information (Figure S3). Additional room-temperature output characteristics and the extracted subthreshold swing are shown in Figures S4 and S5, respectively. The conductivity is typically attributed to $\pi$-electrons in the conjugated macrocyclic ring [25]. This allows charge carriers (holes or electrons) to move within the system. Thus, the electrical properties in vacuum are dominated by the intrinsic bandgap. Without external adsorbates, charge transport is governed by the hopping conduction or band-like conduction [26].

In the absence of a central metal ion, Pcs interact weakly with $N_2$ because the molecules are relatively inert. The interaction between $N_2$ and the $\pi$-system of metal-free phthalocyanine is typically dominated by Van der Waals forces where the $N_2$ molecule can adsorb onto the surface of the Pcs macrocycle via weak, non-covalent interactions. Due to the planar nature of the Pc macrocycle, minimal $\pi$-$\pi$ stacking interactions between the aromatic system of the phthalocyanine and $N_2$ are expected. In some MPcs, the d-orbitals of the metal ion interact with the $\pi^*$ orbitals of $N_2$, which can result in some degree of $N_2$ activation [25], [27], [28]. The interaction may be facilitated by the metal's vacant coordination sites, and $N_2$ can act as a ligand [29], [26]. $H_2Pc$ exhibits the lowest drain current among all investigated devices (Figure S2a), approximately one order of magnitude lower than that of the MPcs. For the MPc-based devices, although the overall transfer-curve shapes differ, the maximum drain current measured at $V_G = -90V$ is comparable across the different metal centers. This indicates that the presence or absence of a metal center significantly affects the electrical response under nitrogen. In vacuum, MPcs are effectively isolated from external molecules, enabling their intrinsic semiconducting properties to dominate the device behavior. As a result, all samples exhibit a substantial increase in current response under vacuum conditions, reflecting the absence of environmental species that typically hinder charge transport [29], [7], [26].

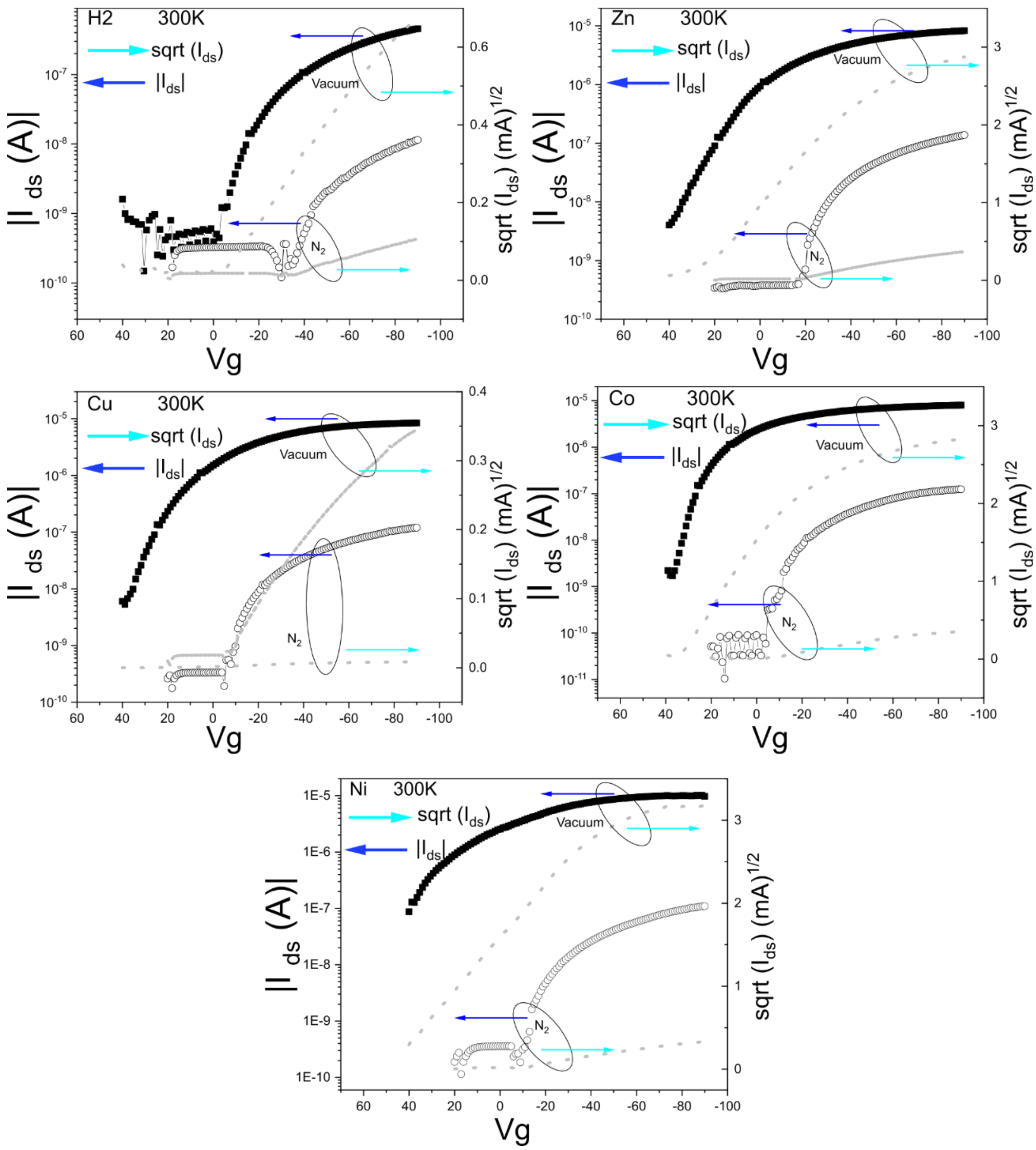

Figure 3. Transfer curves for five P-type semiconductor liquid crystalline compounds ($H_2Pc, ZnPc, CuPc, CoPc, NiPc$), showing a comparison between measurements conducted under vacuum conditions and those carried out in a controlled environment inside the glove box at room temperature (300 K).

The electrical performance metrics of the devices, including mobility ($\mu$), threshold voltage ($V_{th}$), on/off current ratio ($I_{on}/I_{off}$), and the subthreshold swing ($SS$), were derived for each Pc-based OTFT. To evaluate device-to-device reproducibility, electrical measurements under $N_2$ atmosphere were performed on three independent devices for each material, and the mobility values reported in Table 1 correspond to the average values with their respective standard deviations. In addition, the mobility distribution obtained for devices with different channel lengths is presented in Figure S6. For the temperature-dependent measurements, a representative device exhibiting the best electrical performance was selected for characterization in the cryostat system. The mobility values summarized in Table 1 are notably lower than those typically reported in the literature for Pcs [30] [31] [32] [33], [34] [35], [36], which vary from the $10^{-5}$ to the $10^{1}$ cm/Vs range. This reduction in mobility may be attributed to the lateral insulating chains, which confer the liquid crystal properties to the Pcs. Although the current increased under vacuum conditions, the mobility of the devices remained unaffected. The low charge carrier mobility observed in these devices is mainly attributed to the non-ideal orientation of the columnar mesophases relative to the transistor channel. In liquid crystal phthalocyanines, efficient transport requires stacked π–π columns aligned parallel to the source-drain direction; however, in the films in question, the columns are likely randomly oriented or tilted relative to the substrate, leading to interrupted transport paths and limited electronic coupling. The low charge carrier mobility observed in these devices is mainly attributed to the non-ideal orientation of the columnar mesophases relative to the transistor channel. Previous optical and morphological studies of these liquid-crystalline phthalocyanines revealed columnar domains with characteristic dimensions comparable to the 2.5 μm channel length [15,16]. In liquid crystal phthalocyanines, efficient transport requires stacked π–π columns aligned parallel to the source-drain direction; however, in the films in question, the columns are likely randomly oriented or tilted relative to the substrate, leading to interrupted transport paths and limited electronic coupling.

Consequently, charge transport likely involves multiple interdomain boundaries, which can act as energetic barriers and charge-trapping sites. Combined with the random orientation of the columnar mesophases, these effects contribute significantly to the reduced carrier mobility.

These findings indicate that charge transport is structurally limited by columnar orientation, and not by environmental effects or intrinsic electronic structure. A similar trend was observed for the $I_{on}/I_{off}$ ratio. Regarding the threshold voltage, all MPc-based devices exhibit a clear shift toward more negative $V_{th}$ values, whereas $H_2Pc$ shows a shift toward less negative (more positive) threshold voltages. For a p-type transistor, the increase of Vth to more positive values indicates that the device can be turned on with less voltage, which usually corresponds to reduced charge trapping in the device. The SS remained remarkably stable for the $H_2Pc$, but MPcs tested in vacuum revealed a pronounced increase in SS. The SS remained remarkably stable for the $H_2Pc$, but MPCs tested in vacuum revealed a pronounced increase in SS, indicating a significant rise in current under the same applied potential window compared to those measured in $N_2$.

Although vacuum conditions enhance the overall current response, the increased SS values suggest that charge transport remains governed by a broad distribution of localized trap states and energetic disorder. In the absence of adsorbed atmospheric species, intrinsic transport limitations associated with interdomain boundaries and heterogeneous hopping pathways within the liquid-crystalline films may become more pronounced, leading to broader turn-on characteristics.

| | Nitrogen | | | | Vacuum | | | |
|---|---|---|---|---|---|---|---|---|
| Material | $\mu$ | $V_{th}$ | $I_{on}/I_{off}$ | SS | $\mu$ | $V_{th}$ | $I_{on}/I_{off}$ | SS |
| $H_2Pc$ | 0.35±0.11 | 16.8 | $10^2$ | 8.3 | 0.27 | - 4.0 | $10^3$ | 8.17 |
| $ZnPc$ | 2.07±1.12 | 11.5 | $2.2 \times 10^2$ | 2.5 | 4.19 | 28.0 | $27.5 \times 10^3$ | 13.0 |
| $CuPc$ | 3.17±0.99 | 7.4 | $1.7 \times 10^5$ | 2.8 | 4.10 | 35.0 | $5.6 \times 10^4$ | 13.3 |
| $CoPc$ | 5.22±0.73 | 1.6 | $9 \times 10^3$ | 4.4 | 7.51 | 34.3 | $1.7 \times 10^2$ | 5.2 |
| $NiPc$ | 12.07±6.79 | 16.7 | $2.4 \times 10^3$ | 4.4 | 3.62 | 49.0 | $6 \times 10^3$ | 13.7 |

Table 1 Summary of OTFT devices employing metal phthalocyanines (MPcs) as the semiconductor layer. Reported parameters include the field-effect mobility, μ ($\times 10^{-5}\ cm^2\ V^{-1}\ S^{-1}$), threshold voltage, $V_{th}$ ($V$), and subthreshold swing, SS ($V/dec$)

The $V_{th}$ of the OTFT at different temperatures was determined from the plot of square root of the drain current versus gate voltage under VD = − 90 V. Figure 4a shows a shift of $V_{th}$ towards more negative values as the temperature decreases from room temperature to 100 K, which is attributed to the increasing impact of deep traps partially located at the insulator/organic semiconductor interface, a phenomenon that becomes increasingly prominent at lower temperatures. At reduced temperatures, these traps are more effective in capturing charge carriers, thereby affecting the overall device performance [37], [38], [39]. Additionally, charge transport is also limited by bulk trapping, where the movement of charge carriers is impeded by the presence of traps within the bulk of the material. These traps act as localized energy states that capture carriers, leading to a notable reduction in charge transport efficiency [40], [41]. Overall, the temperature dependence of the $V_{th}$ in Pcs is largely due to the decreasing capability of the thermal excitation of carriers to populate deep trap states, thus requiring higher gate voltages to reach the same level of conductivity. Next, we did further electrical measurements at lower temperatures (Figure 4a). All $V_{th}$ would shift towards more negative values upon cooling, and only that of $H_2Pc$ would initially display negative values at room temperature.

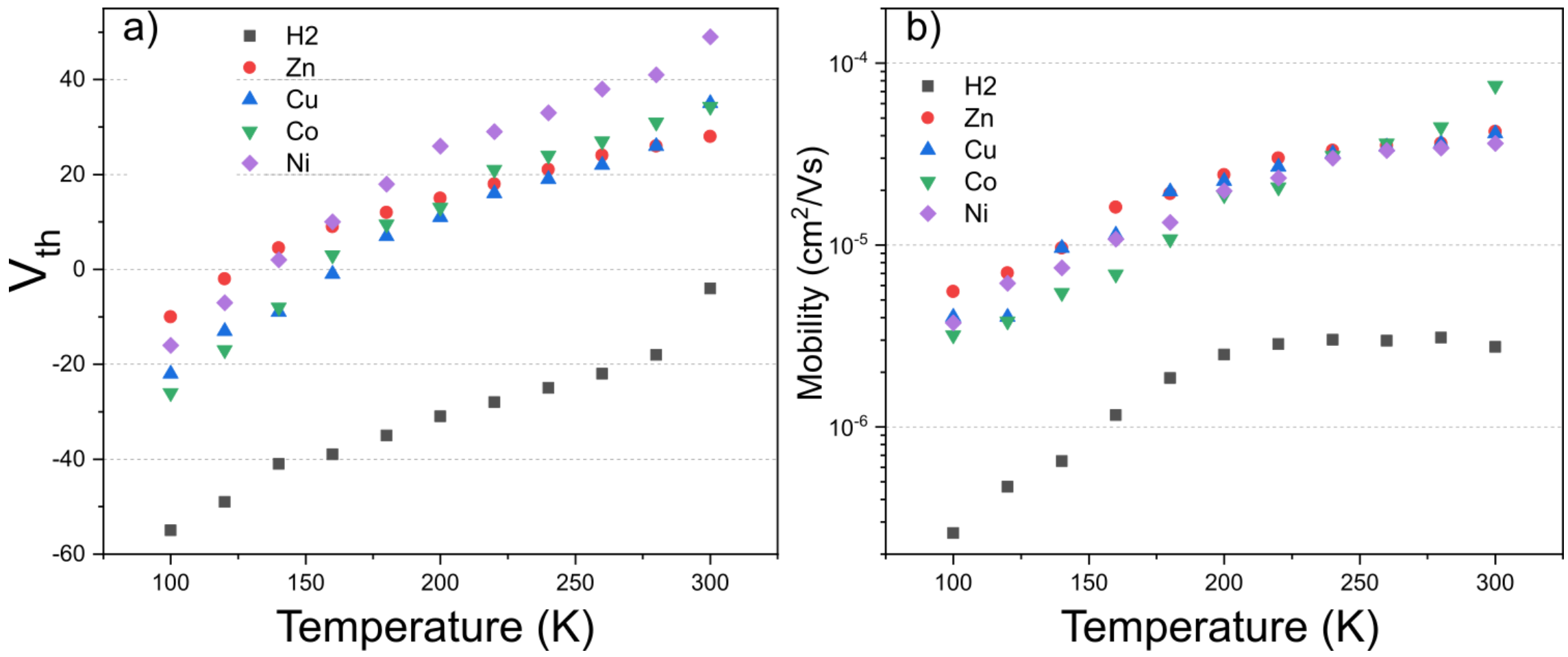

Figure 4. a) Extracted threshold voltage for the five Pc devices. b) Field-effect mobility for the five Pc devices as a function of the temperature.

Temperature dependence provides insight into the charge transport mechanisms within organic semiconductors. The field-effect mobilities for our devices were obtained from the saturation region of the transfer curves by using $\mu = \frac{2L}{WC_i}\left(\frac{\partial\sqrt{I_{ds}}}{\partial V_g}\right)^2$, where $C_i$ is the dielectric capacitance, $W$ is the channel width, $L$ is the channel length, $V_g$ is the gate voltage, $I_{ds}$ is the drain current, and $\mu$ is the field-effect mobility. Figure 4b shows the charge carrier mobility of the Pcs as a function of temperature, which are relatively low in contrast to reported values [26]. This outcome aligns with our expectations, given the insulating nature of the side chains present in the molecules, which might act as trapping sites impeding charge hopping. Within the temperature range of 100 K to 200 K, $H_2Pc$ exhibits a linear increase in mobility, reaching a plateau that persists until room temperature. On the other hand, the mobility values for the MPcs compounds increase by an order of magnitude, demonstrating a consistent linear correlation with temperature.

The decrease in temperature affects the $V_{th}$, *SS* and the on/off ratio. The transition between the on/off states remains relatively stable, attributed to a slight rise in the *SS* (average values calculated in the supplementary material). Initially, the on/off ratio rises until reaching 260 K; however, at lower temperatures, an inversion takes place, causing a sharp decline in the on/off ratio. This change is exclusively commanded by the on state, since the off state basically does not change with temperature (see Figure S7).

For p-type transport, the decrease in conductivity observed upon cooling indicates a reduction in the gate-induced free charge density. The free charge density (N) at the organic–dielectric interface was estimated according to:

$$N = \frac{|V_g - V_{th}|\, C_i}{q},$$

where $V_g$ corresponds to the maximum applied gate voltage ($-90\ V$), $C_i$ is the capacitance of the insulating layer, and $q$ is the elementary charge.

The semiconductor-dielectric interfacial density of traps ($D$) as a function of the temperature offers good insights into the OTFT transport mechanism:

$$D = \left(\frac{q\ SS}{k_B T}\ log(e) - 1\right)\frac{C_i}{q}$$

$SS = \left[\frac{d\log(I_{ds})}{d\,V_g}\right]^{-1}$ and $e \approx\ 2.718$ is Euler's number, $k_B$ is the Boltzmann constant.

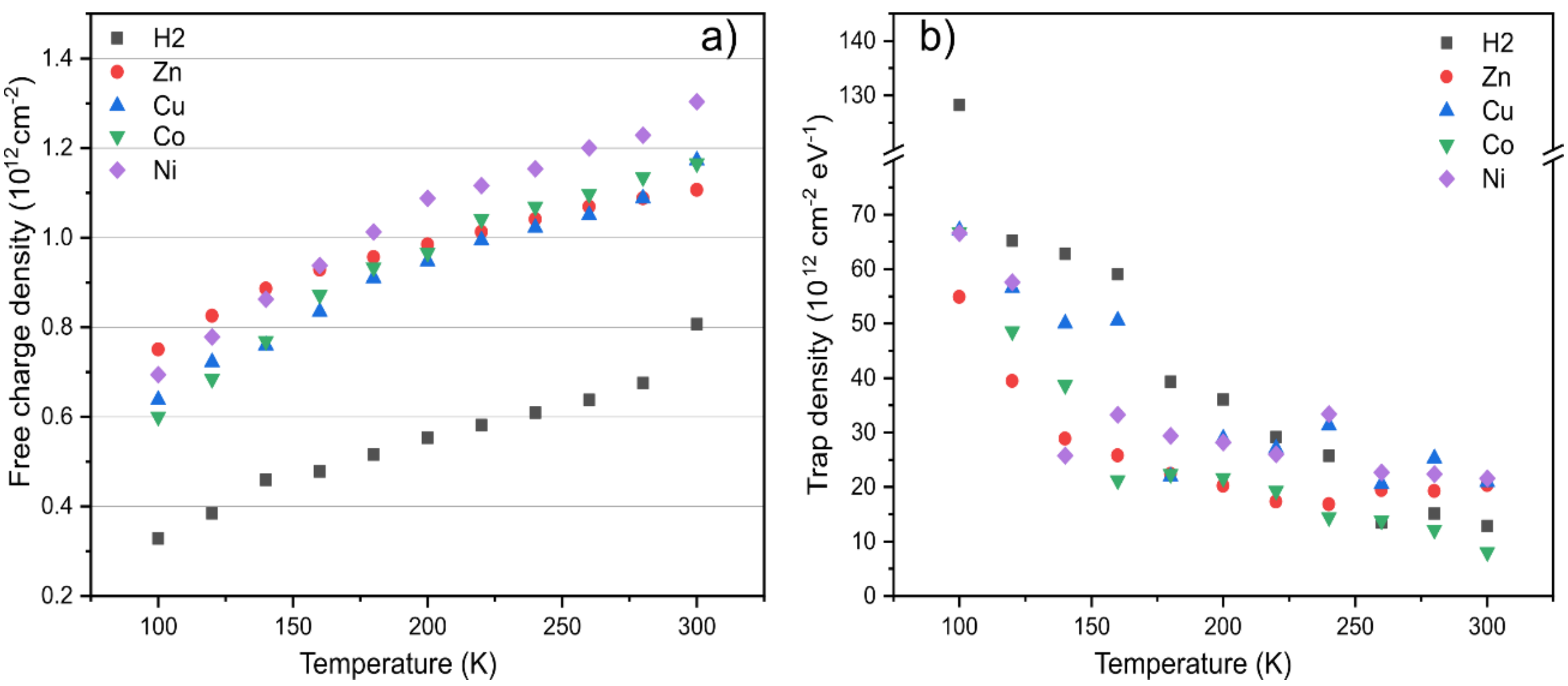

Figure 5 a) Free charge density extracted from the threshold voltage. b) Trap states density extracted from the subthreshold swing.

Overall, the results reported in Figure 5 show that the trapped charge density of all devices increases at lower temperatures, with that of $H_2Pc$ devices being the highest, which is consistent with the findings reported in Figure 4.

The extracted field-effect mobility is plotted in an Arrhenius representation as a function of temperature at $V_G = V_D = -90\ V$ for all devices (Figure 6). The mobilities exhibit thermally activated behavior across the considered temperature range, which is expected in thin organic films [36]. The extracted field-effect mobility of OTFTs follows the simple Arrhenius relation , from which the activation energy ($E_a$) can be extracted by $\mu_{FET} = \mu_0\ exp(-E_a/k_B T)$ [39], with $\mu_0$ being the specific mobility, $k_B$ is the Boltzmann constant, and $T$ is the temperature. The activation energies $E_a$ extracted from the Arrhenius analysis Figure 6, $H_2Pc$ (19.6 meV), ZnPc (13.9 meV), CuPc (17.5 meV), CoPc (15.0 meV), and NiPc (15.3 meV), are uniformly low, consistent with thermally activated hopping in a shallow trap landscape. In liquid-crystalline phthalocyanines, this behavior reflects the presence of local molecular order within mesophases, which minimizes deep trapping, while the dynamic positional disorder and increased intermolecular spacing inherent to LC packing limit electronic coupling and thereby constrain the absolute mobility values [15], [16]. These values fall well below those typically reported for phthalocyanine thin films and other disordered organic semiconductors [43], [44], [45], reflecting the reduced energetic disorder imposed by the columnar liquid-crystalline mesophase. The energy levels of all semiconductor molecules, compiled from thin-film measurements, are summarized in the Supporting Information (Figure S8).

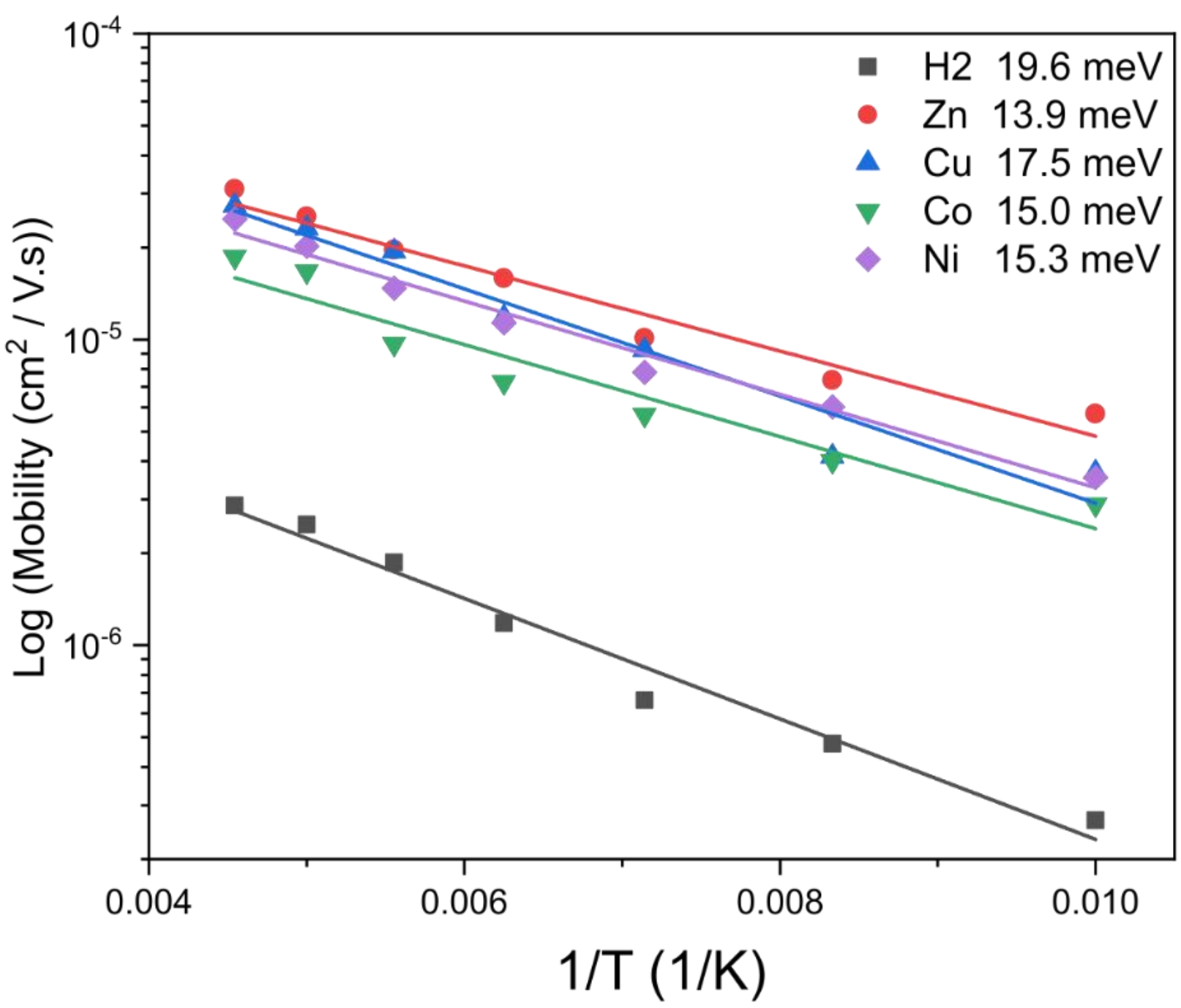

Figure 6 Extraction of the activation energy for the five *p*-type semiconductor liquid crystalline compounds ($H_2Pc, ZnPc, CuPc, CoPc, NiPc$).

## 4. Conclusions

We have demonstrated that liquid-crystalline phthalocyanines ($H_2Pc$ and their Cu, Zn, Ni, and Co complexes) exhibit metal-dependent structural distortions that directly influence their semiconducting behavior in thin-film transistors. Raman analysis revealed systematic shifts in C–N–C and M–N modes arising from differences in metal–ligand coordination, confirming the sensitivity of the macrocycle to the central ion. When incorporated into OTFTs, all compounds showed markedly higher currents under ultrahigh vacuum than in a $N_2$-rich environment, highlighting the strong suppression of intrinsic transport by adsorbed atmospheric species. Temperature-dependent measurements revealed shallow-trap-dominated transport and thermally activated mobilities with low activation energies (≈14–20 meV). These results demonstrate how mesomorphic order, metal coordination, and environmental conditions collectively govern charge transport in liquid-crystalline phthalocyanines. Our investigation therefore constitutes a systematic characterization of a series of liquid-crystalline phthalocyanine OTFTs, revealing low trap-barrier energies and relatively efficient charge-hopping pathways present in these materials.

## Acknowledgments

E.O. acknowledges individual financial support by NSERC (funding Reference No. RGPIN-2023- 0579), the Fonds de recherche du Québec-Nature et technologies (FRQNT), and the Canadian Foundation for Innovation (John R. Evans Leaders Fund, project #: 37358). Z.Y. thanks the Fonds de recherche du Québec for his PBEEE doctoral scholarship. Pieces of this work were funded by CNPq, FAPESC, INCT-INEO, CAPES Foundation, an agency linked to the Ministry of Education of Brazil - PDSE (PDSE - Sandwich Doctorate Program Abroad), number: 88881.690092/2022-01 (Migrated - SICAPES3). The project INCT-INEO FAPESP (project #: 2025/27044-5) is acknowledged for funding.

## Conflicts of Interest

The authors declare no conflicts of interest.